\documentstyle[11pt,newpasp,twoside,epsf]{article}
\def\edcomment#1{\iffalse\marginpar{\raggedright\sl#1\/}\else\relax\fi}
\reversemarginpar

\begin{document}
\title{Quasar Apparent Proper Motion Observed by Geodetic VLBI Networks}
 \author{D. S. MacMillan}
\affil{NVI, Inc., NASA Goddard Space Flight Center, Greenbelt, MD, 20771}

\begin{abstract}
In our standard geodetic VLBI solutions, we estimate the positions of quasars
assuming that their positions do not vary in time. However, in solutions estimating 
proper motion, a significant number of 
quasars show apparent proper motion greater than 50 $\mu$as/yr. For individual 
quasars, there are source structure effects that cause apparent proper motion. To 
examine how coherent the pattern of apparent proper motion is over the sky, we have
estimated the vector spherical harmonic components of the observed proper motion 
using VLBI data from 1980 to 2002. We discuss the physical interpretation of the
estimated harmonic components. 
\end{abstract}

\section{History of Radio Source Observing by Geodetic Networks}
Since 1979, geodetic VLBI networks have observed in 3555 24-hour experiment 
sessions. Most of these observations were made by about 40 antennas. 
The number of radio sources observed in geodetic sessions has grown significantly
during the last two decades. Until 1986, only 65 sources had been observed and these
were nearly all in the Northern Hemisphere. From 1987 to 1989, 210 more sources were 
observed and the distribution of sources between hemispheres was greatly improved.
In our current geodetic solutions, we estimate positions for 610 radio sources. 
Due to the fact that most of the geodetic antennas are in the Northern
hemisphere, the distribution of radio sources is still better in the Northern
than the Southern hemisphere. 

\begin{figure}
\par\vspace{-2mm}
\epsfclipon\plotone{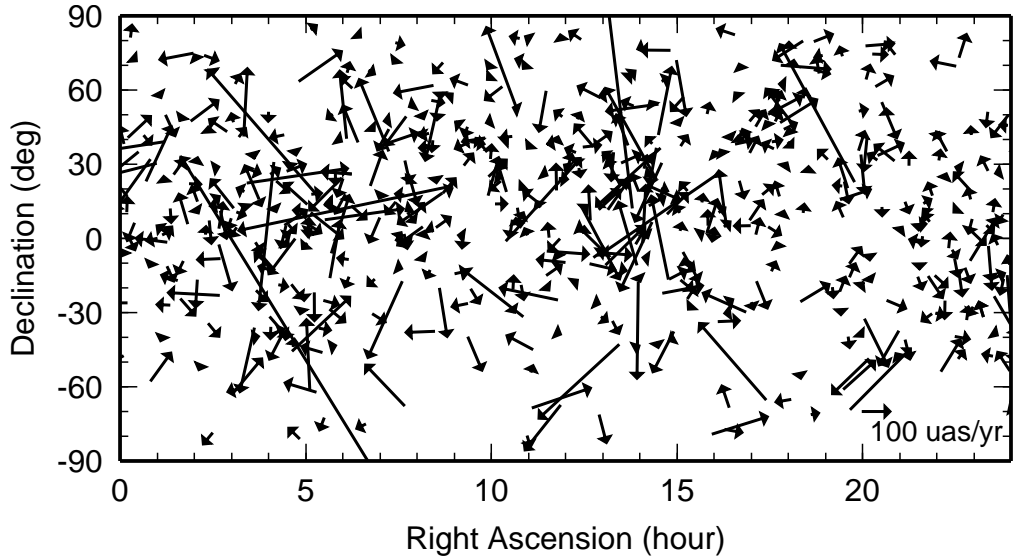}
\caption{Observed magnitude of proper motion for quasars} \label{fig-1}
\par\vspace{-2mm}
\end{figure}

Since 1990, there have been 2-3 24-hour geodetic sessions each week, where
typically 40-50 radio sources are observed per session. Typically each session 
used a network of 4-7 antennas, but we also have data from the bimonthly 
RDV series of sessions (1997-2003) that mostly used 
10 of the standard geodetic antennas along with the 10-station VLBA. In each session,
the choice of radio sources to be observed depended on the geometry of the
network. The result is that the sampling history of the set of all sources 
that have been observed is generally uneven.  
 
\section{Observed Proper Motion Field}
A VLBI terrrestrial reference frame solution was performed using all VLBI observations 
from 1979-2003. In this type of solution, station positions, station velocities, and radio 
source positions and velocities are estimated from all of the data. To remove the translational
and rotation degeneracies of the solution, it is necessary to apply several constraints.
Here, the station positions and velocities 
were weakly constrained to ITRF2000 via no net translation and rotation constraints. 
Similarly, radio source positions were weakly constrained to a priori ICRF positions with 
a no net rotation constraint. Radio source velocities (proper motions) were constrained to
have no net rotation velocity. The paper by Ma et al. (1990) describes the least-squares 
estimation program (SOLVE) used in the analysis. For each 24-hour session observing 
day, the estimated parameters are pole coordinates and their rates, UT1 and the UT1 rate, 
and nutation offsets. Several nuisance parameters are estimated as piecewise linear 
functions with constraints on the rate of change between segments. These parameters are 
station clock functions with one hour segments, wet atmospheric delay with 20 minute 
segments, and horizontal tropospheric delay with 6 hour segments. For the most part, the 
theoretical time delays follow the IERS Conventions (McCarthy 1996). 

Figure~\ref{fig-1} shows the distribution of proper motion for 580 radio sources  
with proper motion formal uncertainties better than 0.5 mas/yr.     
Since the observations of sources have not been very even, the uncertainties of the 
source proper motion estimates range from less than 50 $\mu$as/yr to more than 1 mas/yr. 
The formal uncertainties for 167 sources are better than 20 $\mu$as/yr and better than
50 $\mu$as/yr for 348 sources. There are about 50-60 
sources with observed proper motion with at least 3 sigma significance. In terms of
obvious systematic effects, there is not any clear declination dependence of
observed proper motion in declination or in right ascension. However,
the precision of determinations of proper motion
for southern declination sources (below 40-50\deg S) is poorer than for higher
declination sources. The main reason for this is that most of the geodetic antennas
are in the northern hemisphere.
For the distribution of estimated proper motions, we find that the weighted
RMS of motion in declination was 30 $\mu$as/yr and in right ascension (arc length) was
26 $\mu$as/yr. 

\begin{figure}
\par\vspace{-2mm}
\epsfclipon\plotone{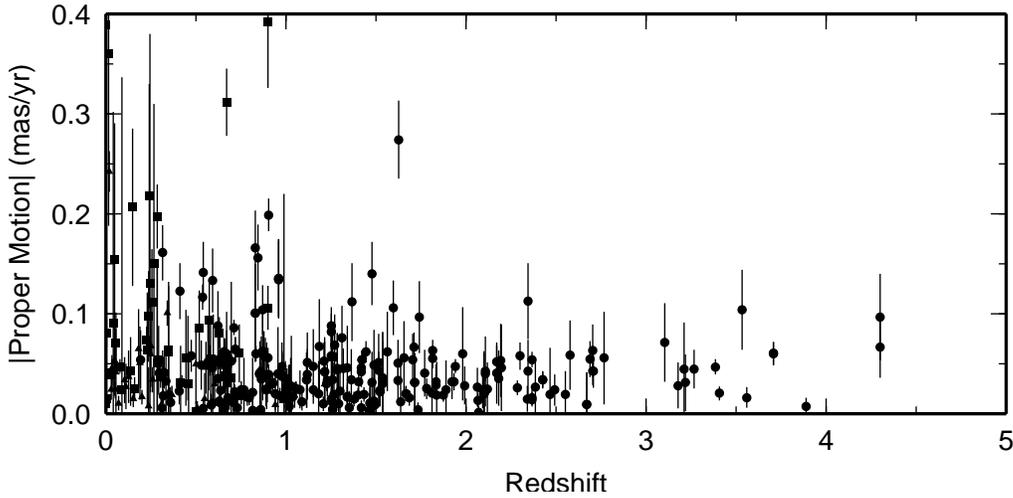}
\caption{Dependence of observed magnitude of proper motion of extragalactic 
radio sources on redshift} \label{fig-2}
\par\vspace{-2mm}
\end{figure}

Figure~\ref{fig-2} shows the magnitude of observed proper motion versus redshift for sources
where the proper motion uncertainty is at least as good as 50 $\mu$as/yr. The available 
redshifts were taken from Archinal (1997), which contains radio source characteristics
compiled from a number of sources. The observed quasars (355 sources) have redshifts
as large as 4.3. The observed radio galaxies (54 sources) and BL Lac objects (56 sources)
generally have redshifts less 
than 1. There may be a trend toward lower proper motion with increasing 
redshift but it is not very clear.

\section{Possible Source Structure Effects}

\begin{figure}
\par\vspace{-2mm}
\epsfclipon\plotone{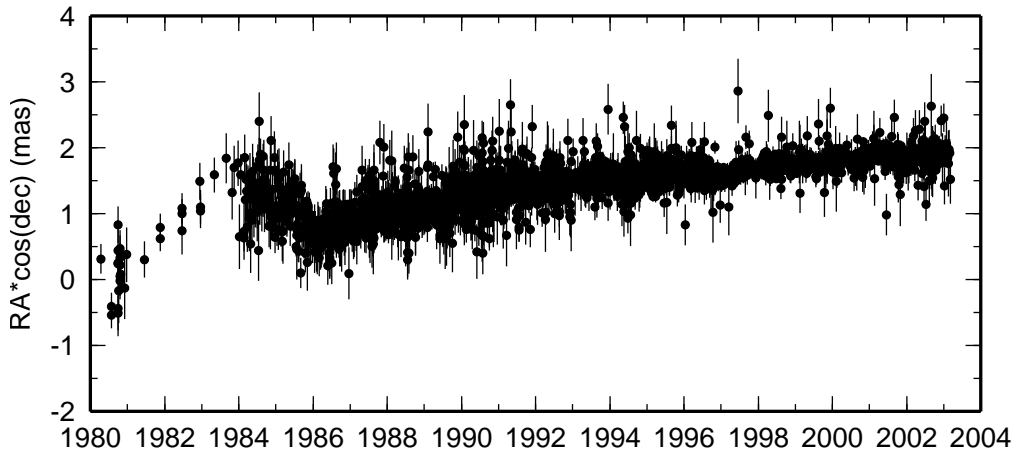}
\caption{Evolution of 4C39.25 position in right ascension}\label{fig-3}
\par\vspace{-2mm}
\end{figure}

Source structure variations are known to be correlated with variations in the apparent
position measured by VLBI (Charlot et al. 1990). When a solution is made in which source
position time series are estimated, one observes both linear and nonlinear variation. 
One problem is that we do not know how much of this observed linear variation is due
to structure variation. Figure~\ref{fig-3} shows the time series for the right ascension of
4C39.25, whose apparent proper motion has been shown to be related to structure 
variations (Fey et al. 1997).
Sovers et al. (2002) have derived structure corrections from
source maps for the RDV series of VLBI sessions and found that these corrections 
removed about 8 ps in quadrature from solution weighted rms delay residuals (typically
at a level of 30 ps).

\section{Pattern of Apparent Proper Motion}

We have analyzed the observed proper motion vector field to determine whether there 
are systematic patterns present. We have expressed the observed field as an expansion
of transverse vector spherical harmonics (VSH), which conveniently comprise an orthonormal
basis for vector fields on a sphere. 

\[ \sum_{l,m} (a^{E}_{l,m} Y^{E}_{l,m} + a^{M}_{l,m} Y^{M}_{l,m}) \] 
 
The E-harmonics are the electric or poloidal harmonics and the M-harmonics are
the magnetic or toroidal harmonics. Since the observed motion is real, we estimate 
real linear combinations of the complex amplitudes. This is similar to the approach
of Gwinn et al. (1997)
except that instead of estimating the expansion amplitudes (externally) from the
set of proper motion estimates made for each source (solution in Section 2), the
expansion amplitudes were estimated directly in the solution. 
We also have available another six years of data.

\begin{figure}
\par\vspace{-2mm}
\epsfclipon\plotone{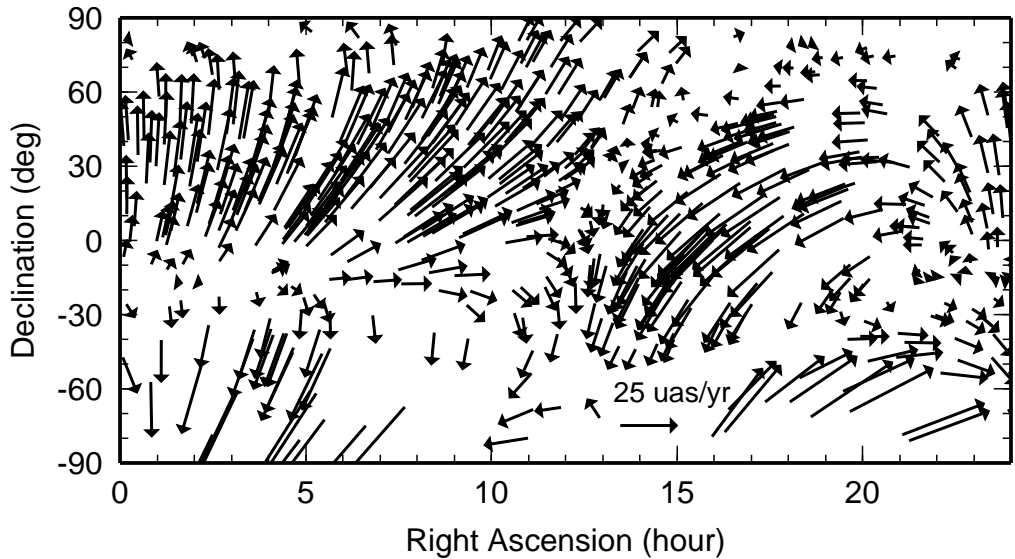}
\caption{L=2 proper motion transverse vector field at the observed source
 positions} \label{fig-4}
\par\vspace{-2mm}
\end{figure}

The L=1 M-harmonics are simply rotations and are therefore indistinguishable from Earth 
rotation. On the other hand, the L=1 E-harmonics can arise from 
galactocentric acceleration or quasar/galaxy acceleration. The RMS of the proper
motion for observed sources using the sum of these harmonics was found to be ~8 $\mu$as/yr. 
For the sum of the L=2 harmonics, we obtained an RMS of 19 $\mu$as/yr. Gwinn et al. (1997)
related the squared proper motion averaged over the sky to the energy density of 
gravitational radiation for wavelengths short compared to source distances. 
Figure~\ref{fig-4} shows the estimated L=2 vector field. 

A variant of the above solution was made in which for each year, the positions for all
sources observed in a given year were estimated with a VSH expansion. One can clearly see
the linear evolution in time of the expansion amplitudes after 1990. For years increasingly 
earlier than 1990, 
the distribution of sources poorer especially in the Southern hemisphere so that 
a spherical harmonic expansion becomes increasingly less reliable.    

\section{Discussion}

The position error floor of the ICRF (International Celestial Reference Frame) is 
250 $\mu$as and the orientation stability is about 20 $\mu$as.
The stability of the celestial reference frame depends on the level of secular change
of radio source positions. The WRMS of observed proper motion from nearly 2 decades of
observations by geodetic networks is at the level of about 30 $\mu$as/yr in both right
ascension and declination.  Given this level of observed motion and an observing
duration of 1-2 decades, modeling or corrections 
are needed to make improvement in stability below the nominal noise floor.

There does appear to be a statistically significant pattern in the observed proper
motion vector field. The problem, however, is to how to determine how much the observed
apparent motion is due to unmodelled source structure effects. To improve the 
determination of spherical expansion amplitudes, it would be desirable to make more
observations of southern declination sources below about -40\deg.


\begin{references}
\reference Archinal, B. A., Arias, E. F., Gontier, A. -M., \& Mercuri-Moreau, C.
 1997, IERS Technical Note 23, III-11 
\reference Charlot, P. 1990, \aj, 99, 1309    
\reference Fey, A., Eubanks T. M., \& Kingham, K. 1997, \aj, 114, 2284
\reference Gwinn, C. R., Eubanks, T. M., Pyne, T., Birkinshaw, M, \& Matsakis, D. N. 1997, \apj, 
485, 87
\reference Ma, C., Sauber, J. M., Bell, L. J., Clark, T. A. ,Gordon, D., Himwich,
 W. E., \& Ryan, J. W. 1990, J. Geophys. Res,
95, 21991    
\reference McCarthy, D. D. 1996, IERS Technical Note 21 
\reference Sovers, O. J., Charlot, P., Fey, A. L. \& Gordon, D. G. 2002, IVS 2002 General
Meeting Proceedings, 243
\end{references}
\end{document}